# Enhancing Data Security Against Cyberattacks in AI-Based Smart-Grid Systems with Crypto-Agility

Marcelo G. Simões[1], Mohammed Elmusrati[1], Tero Vartiainen[1], Mike Mekkanen[1], Mazaher Karimi[1], Sayawu Diaba[1], Emmanuel Anti[1], and Wilson Lopes[2]

(1) University of Vaasa, School of Technology; (2) Thales, Eval Digital

## I. Introduction

Traditional and conventional power systems are significantly impacted by the environmental pollution and low efficiency associated with fossil fuels. These issues have been supporting a new paradigm of electricity generation locally at the distribution level, with renewable and alternative sources, making possible non-conventional distributed energy resources (DERs). Those are typically called microgrids (MGs), there are other denominations as well. The main idea is to have microgrids deployed on low- or medium-voltage active distribution networks. They can be advantageous in many different ways, such as improving the energy efficiency and reliability of the system, reducing transmission losses and network congestion, and integrating of clean energies. Despite those clear advantages, there are yet challenges in implementing MGs with DER units, those are related to power quality and stability issues – MG's voltage and fault level changes, energy management, low inertia, further complex protection schemes, load and generation forecasting, cyber-attacks, and cyber security. MGs should operate in grid-connected and isolated modes, with energy management and protection schemes becoming intricate than those in the usually distributed networks. Moreover, due to the rapid load variation in MGs and the variable renewable energy resource generation, load/generation forecasting is needed in applications such as energy management. Microgrids rely on information and communication technologies, and as such, their security is critical, and they might be vulnerable to different types of cyber-attacks so cybersecurity techniques can provide safe operation of MGs. In the past few years, MG evolved in Smart-Grids, as discussed in this paper. In order to address all those challenging features, this paper shows the deep utilization of advanced, accurate, and fast methodologies such as artificial intelligence (AI)-based techniques. They guarantee efficient, optimal, safe, and reliable operation of smart grids safe against cyberattacks. AI refers to the computer-based systems' ability to perform tasks with intelligence typically associated to human decision-making. AI-based systems can learn from past experiences and solve problems. AI has been used in different applications, including MGs, to improve system performance.

## II. Future Digital Protection System Challenges and Issues

Power system protection refers to the measures taken to detect and isolate faults in the power system to prevent damage to equipment, avoid power outages, and protect people and property from potential hazards. This protection system involves the use of various components such as relays, circuit breakers, fuses, and other protective devices that act as a defense mechanism against faults in the system. Moving forward with new technologies and digitalization, the future digital protection systems refer to the latest technology being used to protect power systems and their associated equipment [1] with the utilization of intelligent electronic devices (IEDs). On the other hand, digital protection systems rely on communication networks and computer-based algorithms to identify faults and disturbances in the power system and provide automated responses to mitigate the effects of these issues. In fact, power system protection and security are critical issues that must be addressed to ensure the safe and reliable operation of the power grid [2].

The modernization of power grids and the adoption of digital technology have led to significant advancements in the automation, monitoring, and control of power systems [3, 4]. Substation automation systems (SAS) are a collection of intelligent electronic devices (IEDs), communication networks, and computer software that work together to automate the management and control of electric power substations [5, 6]. These systems help in monitoring, controlling, and protecting the substation equipment, and ensure the reliable and efficient operation of the power grid. It is also clear that SAS and future digital protection systems have emerged as critical components of modern power grids, playing a crucial role in ensuring the efficient and reliable operation of the power system in the future infrastructure. However, with the increasing reliance on digital technology, these systems are becoming more vulnerable to cyber-attacks, posing a significant challenge to the security of the power grid [7].

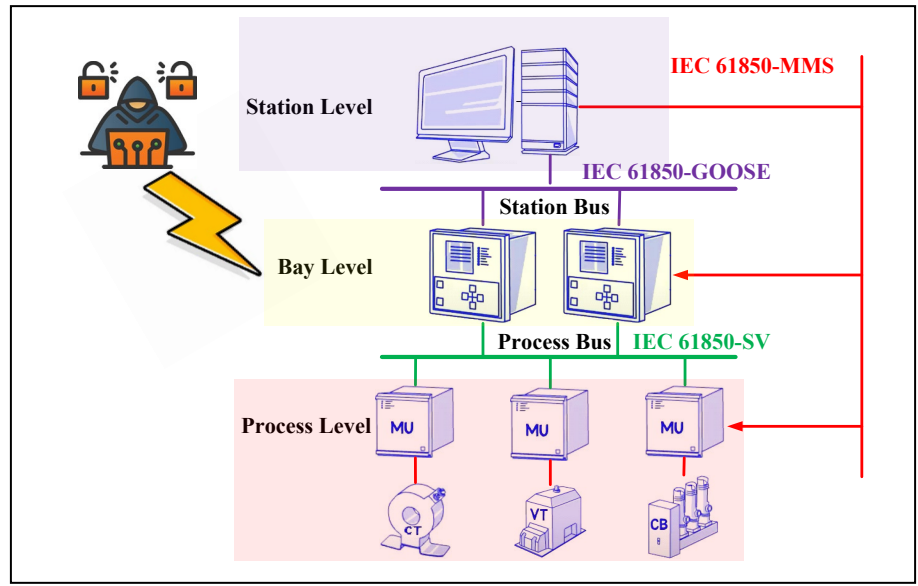

In addition, security risks and vulnerabilities are arisen due to communication infrastructure and interconnected systems [8-10]. Some of the key cyber security issues associated with SAS and digital protection systems are highlighted below:

1. **Cyber-attacks**: One of the significant cyber security issues associated with SAS and digital protection systems is the risk of unauthorized access. Cyber-attacks can disrupt the normal operation of SAS and digital protection systems. These systems are connected to the Internet, making them vulnerable to hacking and other cyber-attacks. Attackers can exploit vulnerabilities in the communication networks, software, and hardware components of these systems to gain unauthorized access, manipulate critical parameters, disrupt power flow, or cause physical damage to equipment.
2. **Malware**: Malware is a type of software that is designed to harm computer systems. It can be introduced into the substation automation system or digital protection system through phishing

emails or other means, leading to the disruption of the system's operation and potentially causing physical damage. Malware attacks can infect the SAS and digital protection systems and disrupt their operation. Malware can cause significant damage to the systems.
3. **Phishing attacks**: Phishing attacks are a common form of cyber-attack that target the users of SAS and digital protection systems. These attacks involve sending deceptive emails or messages that trick users into revealing sensitive information such as passwords or other confidential data.
4. **Lack of proper security protocols**: Sometimes, the lack of proper security protocols can also leave the SAS and digital protection systems vulnerable to cyber-attacks. This may include weak passwords, unsecured communication networks, or outdated software that can be exploited by attackers. Moreover, different security protocols, make it difficult to maintain a consistent security posture. With the introduction of new devices and software, it is essential to ensure that they are compatible with the existing security protocols to avoid potential vulnerabilities.
5. **Insufficient cybersecurity awareness**: Insufficient cybersecurity awareness among the users of SAS and digital protection systems can also contribute to cybersecurity issues. Employees and stakeholders who are not aware of potential cyber threats may inadvertently create security vulnerabilities or fail to report suspicious activity.

To address these challenges, power system operators need to implement comprehensive cyber security measures that include regular vulnerability assessments, employee training programs, and network segmentation. It is also important to ensure that all devices and software used in the system are regularly updated and patched to prevent security breaches. Furthermore, substation automation systems and digital protection systems must be designed with security in mind. Security measures such as access controls, authentication, and encryption must be integrated into the system's architecture to prevent unauthorized access and data breaches.

## III.   ARTIFICIAL INTELLIGENCE AND MACHINE LEARNING

In the past few years there has been an overwhelming enhancement of artificial intelligence (AI) and machine learning (ML) based applications in scientific and engineering fields. These technologies have been branching and penetrating all fronts, from governmental and military operations, to financial, security, medical, educational, industrial, business, and even the most mundane entertainment of music, arts, movies, and the huge streaming industry.

AI and ML go back to their origins around the $2^{nd}$ World War, and there has been at least three waves, the knowledge develops, is applied, hits a constraint in the application, there is an ebb and flow, cycling, and rebirth. In the past few years, there has been another momentum of AI and ML, which became fuelled with deep learning techniques for massive neural networks, with advances and developments in computing chips, wireless communication, powerful compiled and scripted languages, with efficient graphic user interfaces that allow even first year computer scientist students to implement very sophisticated solutions, in a few minutes of work, while 20 years ago any algorithm with such complexity would take months to design, implement, debug and deploy.

Machine learning refers to the techniques that are used to extract information from data. The quality of machine learning algorithms depends on the quality of available data. There are many different techniques for machine learning, and each of them is based on certain assumptions and influential in specific applications. However, typically it is possible to classify machine learning algorithms into four classes: supervised, semi-supervised, unsupervised, and reinforcement learning algorithms. The term AI is more comprehensive, and very often ML is considered a subset (although very strong) of AI. It is also possible to have Intelligent Agents (IA), as an evolution of the original MAS – Multi-Agent Systems, to allow algorithms for proper decisions based on the extracted information from data provided by the ML algorithm.

In Electrical Power Systems, there are many layers of controllability and several fronts of multidisciplinary integrated cooperative work of electrical engineering, computer science, computer technology, and control systems, and a critical application of AI and ML is in the area of cybersecurity for Smart-Grid applications.

## IV. SMART-GRID

Since 2007 there has been a terminology adopted for microgrids with enhanced control, intelligence and capable of interacting with the local utility and be autonomous as well. Smart-grid technology became initially a fever, a buzzword, and currently is accepted as a good terminology of how power systems changed their traditional unidirectional power-flow, to a more intelligent and inter-digital paradigm shift, allowing bidirectional power-flow, as well as ancillary services, net-metering and integration of renewable energy, storage solutions, and electric vehicles integration. Such a transition transformed the ancient power grid, by incorporating cutting-edge advanced communication, control, and monitoring technologies, at the distribution level. Cutting-edge solutions and technologies have been widely used in digital power systems (smart-grid) to optimize and improve the efficiency of electricity generation, distribution, and consumption. There is a simple definition of smart-grid: an integrated dialogue of electrical power flow with signal communications flow in real-time. The figure illustrates how the original microgrid, associated to a power electronics enabled power system, with Artificial Intelligence, plus implementation of Smart Solutions, gets transcended to the status of a Smart-Grid implementation.

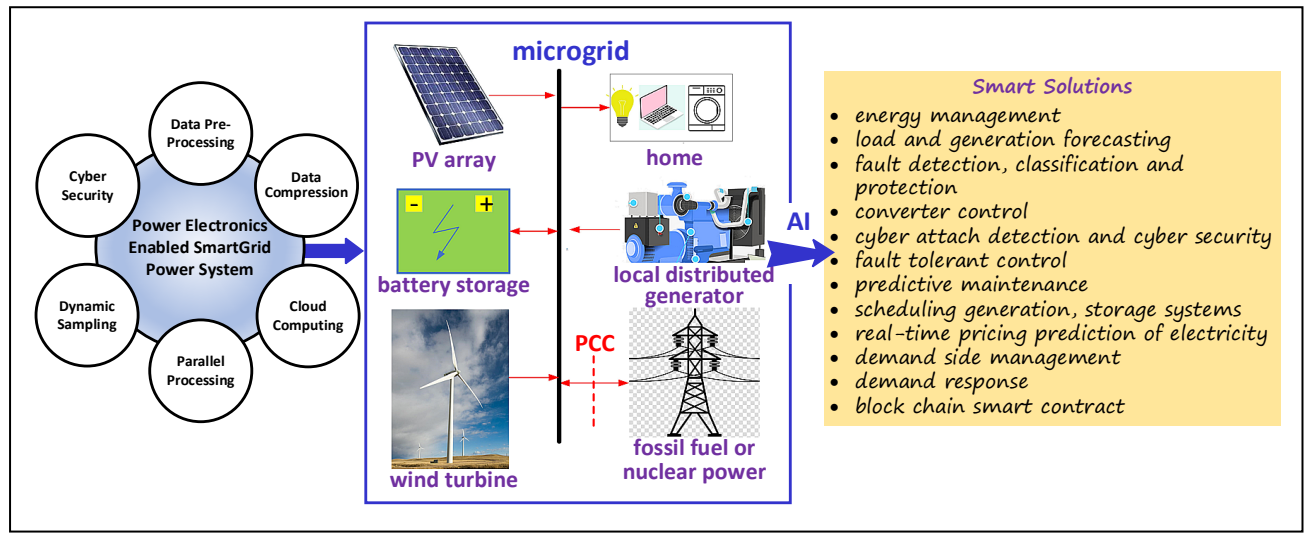

The ability of smart grids to enable bidirectional communication between entities is one of their key features. This enables the implementation of more advanced monitoring, control, and protection functionality, which leads to better energy demand management and allows prosumers to participate in demand response programs. Furthermore, smart grid intelligently facilitates the integration of distributed energy sources (DER) into the grid, assisting in the reduction of carbon emissions and the mitigation of climate change.

## V.    IT AND OT SYSTEMS

The smart grid creates complex systems by combining IT (Information Technology) and OT (Operational Technology), both of which are critical components of smart grid adoption. IT systems provide the communication and data computing capabilities required for smart grid efficiency in which that need to collect and analyze a massive amount of data generated by the grid. Physical infrastructure, such as power generation, transmission, and distribution, is managed by OT systems.

Furthermore, the integration of IT and OT systems in smart grids can result in the emergence of new cybersecurity risks, and because availability from the initial security triangle is the highest priority, malicious actors may exploit vulnerabilities in the IT/OT systems to disrupt grid operations and potentially harm the physical infrastructure. Utilities must employ a multi-layered strategy to mitigate these risks while improving smart grid resilience. This strategy includes technical robust cybersecurity measures as well as policies, procedures, best practices and ensure that employees are well trained to fill skill gaps and follow specific industry requirements in order to raise their awareness and improve their ability to identify and respond to the increasingly raised of the sophisticated security threats.

## VI.    AI AND ML BASED CYBERSECURITY OF SMART-GRIDS

Smart grids are a part of critical infrastructures of modern societies. Critical infrastructures are the systems and assets required for a society's and economy's operation (Boyd et al., 2020; Curt & Tacnet, 2018). These systems include physical infrastructure, such as telecommunications, electric power systems, banking and finance, transportation, water supply systems, cyberinfrastructure, information technology networks, and data centers. People and businesses rely on critical infrastructure daily, and any disruption or damage to these systems can seriously affect public safety and national security. Plachkinova and Vo (2023) created a taxonomy tool for risk assessment of critical infrastructure. The taxonomy is divided into three dimensions: assets, risk management, and attacker motivation. Assets can be cyber, cyber-physical, and physical. Risk Management concerns threats, vulnerabilities, and controls. Attacker motivations can be socio-cultural, political, economic, and thrill-seeking. The tool provides a unified framework to assess cyberattacks on critical infrastructure. In the case of smart grids, valuable assets such as power plants and micro-grids are under threat. The power plants evaluate their threats and vulnerabilities and attempt to prevent attacks motivated by political or economic interests.

The energy sector provides power and fuel to all other sectors, such as communication, transportation, and water (CISA, 2019). In turn, the energy sector relies on transportation for fuel delivery, electricity generation, and infrastructure control and operation through communication. Because of the interdependencies between these infrastructure sectors, the loss of one or more lifeline functions impacts the operation or mission in several sectors. For example, during natural disasters, water and wastewater systems cannot provide clean water until the energy system is restored. The connections explain why critical infrastructures such as smart grids must be protected from physical and cyber-attacks.

Technology is advancing at an unprecedented pace, at the same time there is an ever-increasing need for cybersecurity measures, especially when it comes to protecting critical infrastructure, such as utility-oriented T&D as well as Smart-Grid, particularly because prosumer-based solutions can become a target of hackers as well. Integrating ML and AI in cybersecurity has completely revolutionized how it is possible to approach security measures. It is paramount to apply cutting-edge technologies in cybersecurity to stay ahead of hacktivists. Despite the potential benefits, there are also some drawbacks, as we also delve into the risks, challenges, and threats associated with implementing AI in smart grid security. Professionals today need to have a comprehensive understanding of the critical role that ML and AI play in securing the smart grid.

AI algorithms are capable of analysing vast amounts of data in real-time and identifying patterns that may indicate a security threat. They can be used in cybersecurity for various purposes, such as intrusion detection, malware detection, and threat intelligence. The use of ML and AI in intrusion detection can help to identify abnormal behaviour and malicious activities in a network. Similarly, AI-based malware detection systems can identify new and unknown malware by analysing their behaviour. Moreover, ML and AI techniques can be used to develop threat intelligence systems that can automatically detect and mitigate new and emerging threats.

However, the implementation of ML and AI in cybersecurity also poses specific challenges, such as the need for large amounts of high-quality data, the difficulty of interpreting AI outputs when using black-box learning tools such as deep learning, and the potential for attackers to manipulate AI models. One of the significant risks is the potential for cyber-attacks on the AI systems, which could compromise the entire smart grid security network. Additionally, there is a challenge in ensuring that the AI systems are designed with the necessary security features to prevent unauthorized access and hacking. Another risk is the potential for errors in the AI algorithms, which could lead to false alarms or missed detections. Furthermore, the complexity of the smart grid security system can also pose a challenge in integrating and managing AI systems effectively. These risks and challenges could threaten the reliability and security of the smart grid system, which could have significant consequences. Therefore, it is crucial to address these risks and challenges by developing robust AI systems that are designed with the necessary security features and implementing effective strategies to manage and monitor the AI systems to prevent potential threats and attacks.

Despite these challenges, the use of ML and AI in cybersecurity is expected to continue to grow as organizations seek to improve their detection and response capabilities in the face of an ever-evolving threat landscape. Thus, AI would play a vital role in enhancing the cybersecurity of smart grids by detecting and mitigating potential threats in real-time. Also, AI can help improve the resilience of smart grids by providing intelligent decision-making capabilities during a cyber-attack. One potential future direction for AI in smart grid cybersecurity is the development of AI-powered intrusion detection systems that can learn from historical data and identify new attack patterns. Another opportunity for AI in smart grid cybersecurity is the use of ML algorithms to predict potential cyber-attacks and develop proactive measures to prevent them. Also, AI can be used to analyse the vast amounts of data generated by smart grids to identify anomalies and potential security breaches. But, adopting AI in smart grid cybersecurity also raises ethical and privacy concerns that must be addressed. As such, future research should focus on developing AI-based cybersecurity solutions that are transparent, explainable, and comply with privacy regulations.

General Applications of AI and ML in cybersecurity for smart grids include threat detection and prevention: AI and ML can be used to detect potential cyber threats and prevent them from causing any damage. They can analyse large volumes of data, identify patterns, and detect anomalies that may indicate a cyber-attack. Intrusion detection and prevention: AI and ML can also be used to detect and prevent unauthorized access to the smart grid. They can analyze network traffic and detect any suspicious activity, such as unauthorized login attempts or unusual data transfers. Malware detection: AI and ML can be used to detect and prevent malware from infecting the smart grid. They can analyse software behaviour and detect any malicious activity that may indicate a malware attack. Risk assessment: AI and ML can be used to assess the risk of a cyber-attack on the smart grid. They can analyse various factors, such as the type of devices connected to the grid, the software used, and the potential vulnerabilities. Incident response: AI and ML can be used to automate incident response in case of a cyber-attack. They can detect the attack, analyse the extent of the damage, and take appropriate action to mitigate and reduce the damage consequences.

| CHALLENGES | RISKS |
|---|---|
| Smart grids are complex systems, and integrating AI and ML in cybersecurity requires a thorough understanding of the entire system. The effectiveness of AI and ML algorithms depends on the data quality. Inadequate, inaccurate, or biased data can lead to incorrect conclusions and inadequate security measures. | Using AI and ML in smart grid cybersecurity also poses several threats. One of the most significant threats is data breaches, where attackers can steal sensitive data and use it to launch a cyber-attack. Additionally, cyber-attacks targeting AI and ML systems can exploit vulnerabilities and cause serious harm. |
| There is a shortage of cybersecurity professionals with the necessary skills and knowledge to work with AI and ML, exacerbating the challenge of securing smart grids. | Insider threats are also a concern, where employees with access to the system can misuse the data or manipulate the algorithms to cause harm. This risk is particularly challenging to address, as insiders may have legitimate access to the system, making it challenging to identify and prevent malicious activity. It is essential to implement appropriate access controls and monitoring mechanisms to minimize this risk. |
| Attackers can manipulate the data to deceive the algorithm and cause a cyber-attack. This vulnerability could result in false alarms or missed cyber threats, leading to severe consequences for the smart grid. | |
| Legal and ethical concerns of AI and ML in cybersecurity, including data privacy and the potential algorithmic bias must be addressed. | |

| OPPORTUNITIES |
|---|
| Improved Cybersecurity: AI and ML can provide advanced cybersecurity measures that can protect smart grids from cyber-attacks and ensure the continuity of the energy supply. |
| Automation: AI and ML can automate many cybersecurity processes, reducing the workload of cybersecurity professionals and allowing them to focus on more critical tasks. |
| Predictive Analytics: AI and ML can provide predictive analytics, allowing organizations to predict potential cyber threats and take appropriate measures to prevent them. |
| Efficiency: AI and ML can increase the efficiency of cybersecurity measures, reducing the time and resources required to manage and maintain the system. |

## VII. CYBERATTACKS AND PROTECTION TO ENHANCE RESILIENCY IN SMART-GRIDS

Cyberattacks can take many forms, including malware attacks, denial-of-service attacks, and phishing attacks, and how they can affect the smart grid availability, integrity, and confidentiality. Cyberattacks become more sophisticated, and achieving a secure smart grid is not an easy task. Cyberattacks can target both the IT systems and the OT systems, but the impact is more apparent the on OT systems. Cyberattacks can cause disrupt the supply of electricity and can have serious consequences, including power outages, significant equipment damage, and data losses. To address these challenges, mitigate these risks, and develop effective cyber security measures. Utilities must implement a multi-layered approach to cybersecurity. In this regard, the European Union Agency for Cybersecurity (ENISA) contributes [2012] five pillars: "preparedness, prevention, detection, response, mitigation, and recovery", also National Institute of Standards and Technology (NIST) [2014] proposes five pillars: "Identify, protect, detect, respond and recover", whereas IEC System Committee – Smart Energy (SyC-SE) [2019] provide five Key Cybersecurity Concepts applicable to Cyber-Physical Systems security which are "Resilience, IT versus OT, standard and guides, risk assessment and security by design ". This includes technical controls, policies, and procedures to ensure that employees are trained to identify and cope with the reality of deliberate cyber-attacks and natural disasters, as well as how to maintain/enhance smart grid resiliency and ensure the business continuity.

However, smart grid resiliency measurements/enhancements face unique challenges in determining their feasibility/suitability in the context of smart grid interdependence and system complexity. As a result, smart grid operators must have in-depth knowledge of their systems, dynamic capacities, and capabilities. This deep knowledge must be evaluated on an ongoing basis, along with different dimensions perspectives, in order to draw lessons on extracting an essence into useful entities for resiliency evaluation metrics. Thus, in order to evaluate smart grid resiliency and produce consistent resiliency measurements, different resiliency dimensions must be included alongside the continuous evolution process, and several outcome indicators must be tracked [4,5].

## VIII. CONCLUSION

The integration of AI and ML in the security of smart grids has become a necessity for protecting critical infrastructure from potential cyber-attacks. Although there are risks and challenges associated with the implementation of these technologies, the benefits outweigh the risks. AI and ML can help detect and prevent cyber-attacks, mitigate risks, and improve the overall security of smart grids. Furthermore, using AI and ML in cybersecurity presents opportunities for innovation, increased efficiency, and cost reduction. As the smart grid continues to develop and expand, it is essential to continue to explore and invest in these technologies to ensure the safety and security of the smart grids.